\documentclass[12pt, a4paper]{article}
\usepackage{pstricks}

\usepackage{tikz}
\usetikzlibrary{matrix,arrows,decorations.pathmorphing}

\usetikzlibrary{shapes,arrows}
\usepackage{pgf}
\usepgfmodule{shapes}
\usepgfmodule{plot}
\usetikzlibrary{decorations}
\usetikzlibrary{arrows}
\usetikzlibrary{snakes}

\tikzstyle{decision} = [diamond, draw,
    text width=4.5em, text badly centered, node distance=3cm, inner sep=0pt]
\tikzstyle{block} = [rectangle, draw,
    text width=5em, text centered, rounded corners, minimum height=4em]
\tikzstyle{line} = [draw, -latex']
\tikzstyle{cloud} = [draw, ellipse]

\usepackage{latexsym}
\usepackage{amsmath, amscd}
\usepackage{amssymb}
\usepackage{graphicx}
\usepackage{euscript}
\usepackage{amsfonts}
\usepackage{verbatim}
\usepackage{epsfig}
\usepackage[margin=2cm,bottom=2.0cm]{geometry}
\usepackage{hyperref}

\def\d {{\rm d}}

\def\0         {{\it 0}}
\def\1         {{\it 1}}
\def\2         {{\it 2}}
\def\3         {{\it 3}}
\def\4         {{\it 4}}
\def\5         {{\it 5}}
\def\6         {{\it 6}}
\def\7         {{\it 7}}
\def\8         {{\it 8}}
\def\9         {{\it 9}}

\def\calf         {{\cal F}}

\def\calk         {{\cal K}}
\def\call         {{\cal L}}

\def\calo         {{\cal O}}

\def\cals         {{\cal S}}

\def\rhdot         {r.h.}
\def\lhdot          {l.h.}

\def\del          {\partial}
\def\delbar       {\bar\partial}
\def\ii           {{\rm i}}

\begin{document}

\begin{titlepage}
\titlepage
\begin{flushright}{\tiny ROM2F/2012/02\\
CERN-PH-TH/2012-050}
\end{flushright}
\vskip 2.5cm
\centerline{ \bf \huge Freezing E3-brane instantons with fluxes}
\vskip 2.3cm

\begin{center}
{\bf \large Massimo Bianchi$^1$, Andr\'es Collinucci$^{2, 3}$ and Luca Martucci$^1$}
\vskip 0.3cm
\em
{\it$^1$ I.N.F.N. Sezione di Roma ÒTorVergataÓ \& Dipartimento di Fisica, Universit\`a di Roma ÒTorVergataÓ, Via della Ricerca Scientica, 00133 Roma, Italy}
\\[.5cm]
{\it $^2$ Theory Group, Physics Department, CERN CH-1211 Geneva 23, Switzerland}
\\[.5cm]
{\it $^3$ Physique Th\'eorique et Math\'ematique
Universit\'e Libre de Bruxelles, C.P. 231, 1050 Bruxelles, Belgium}
\vskip 1.2cm

\vskip 2.5cm

\large \bf Abstract
\end{center}

E3-instantons that generate non-perturbative superpotentials in IIB $\mathcal{N}=1$ compactifications are more frequent than currently believed. Worldvolume fluxes will typically lift the E3-brane geometric moduli and their fermionic superpartners, leaving  only the two required universal fermionic zero-modes.  We consistently incorporate $SL(2, \mathbb{Z})$ monodromies and world-volume fluxes in  the effective theory of the E3-brane fermions and study the resulting zero-mode spectrum, highlighting the relation between F-theory and perturbative IIB results. This leads us to a IIB derivation of the index for generation of superpotential terms, which reproduces and generalizes available results. Furthermore, we show how worldvolume fluxes can be explicitly constructed in a one-modulus compactification, such that an  E3-instanton has exactly two fermonic zero-modes. This construction is readily applicable to numerous scenarios.

\vskip2cm

\vskip1.5\baselineskip

\vfill
\vskip 2.mm
\noindent {\small \it Based on a talk by L.~M.~at the XVII European Workshop on String Theory 2011, Padova, Italy, 5-9 September 2011.
\noindent }
\end{titlepage}

\tableofcontents

\section{Introduction}
The appearance of the KKLT scenario \cite{Kachru:2003aw}, which requires a superpotential generated by non-perturbative effects such as E3-brane instantons or gaugino condensates, has sparked a massive search for compactifications that geometrically allow for such effects. In F/M-theory language, a popular approach is to build Calabi-Yau fourfolds that accommodate divisors with holomorphic Euler characteristic  equal to one, $\sum_{k=0}^3(-)^k h^{0,k}=1$, in order to satisfy Witten's weak criterion \cite{Witten:1996bn}. One may also try to satisfy the strong criterion, seeking for {\em rigid} divisors, i.e.\ with $h^{0,3}=0$, and in addition $h^{0,1}=h^{0,2}=0$. Such divisors would support an M5-brane instanton.
Examples of systematic searches for such manifolds are \cite{Denef:2004dm}.

The problem can be translated into IIB language, where the sought-for object is a Calabi-Yau threefold orientifold setup containing a {\em rigid} four-cycle ($h^{0,2}=0$) with  $h^{0,1}=0$ and such that the cycle is orientifold invariant but transverse to the O7-plane. Such a divisor would support an E3-brane instanton with exactly two fermionic zero-modes.

It is known that turning on 3-form fluxes can lift some of the unwanted zero-modes of E3-brane instantons \cite{Hflux}. However, constructing such fluxes explicitly in concrete CY threefolds is not practical. On the other hand, the E3-instanton itself supports a gauge-field whose non-trivial flux can lift the geometric moduli of an otherwise non-rigid E3-brane, such that it effectively becomes rigid. This mechanism can be encoded in a brane superpotential \cite{lucasup}. In fact, that such fluxes can freeze D-branes wrapped on divisors has been exploited in papers that study black hole microstate counting \cite{Gaiotto:2006wm}. However, the technology has not made its way into the instanton literature. Part of the reason might be that one expects such fluxes to be intractable or very difficult to construct. We will show that this not necessarily the case, discussing a case in which it is very simple and intuitive to find fluxes that can entirely freeze an E3-brane.
This note summarizes (some of) the main results of \cite{Bianchi:2011qh}, to which we refer more details.

\vskip 3mm

\section{The fluxless case}
F-theory vacua are type IIB backgrounds that include fully back-reacting 7-branes, whose presence is signaled by a non-trivial axion-dilaton $\tau$, required by supersymmetry to depend holomorphically on the internal coordinates of the compactification manifold $X$:
\begin{equation}
\bar \partial \tau = 0\,.
\end{equation}
There can be either D7-branes or more general $(p,q)$ 7-branes,
obtained by acting on a D7-brane -- i.e.\ a (1,0) 7-brane -- by an SL(2,$Z$) duality transformation.
A D7-brane sources one unit of Ramond-Ramond (RR) flux $F_{ 1 }=\d C_{\it 0}$ and is then characterized by a monodromy $\tau\rightarrow \tau+1$ of the axion-dilaton $\tau:=C_0+\ii e^{-\phi}$ on a closed loop linking the D7-brane. Analogously, a $(p,q)$ 7-brane is characterized by a more general  SL(2,$\mathbb{Z}$) monodromy of $\tau$:
\begin{equation}
\label{dualtau}
\tau \quad \rightarrow\quad \tau^\prime=\frac{a\tau+b}{c\tau+d}\,, \quad \text{with $a,b,c,d\in\mathbb{Z}$ and $ad-bc=1$.}
\end{equation}

The idea behind F-theory is to encode the data of the space $X$ and the axio-dilaton $\tau$ together into a single {\it torus-fibered} manifold $Y$. The latter is defined by a projection $\pi: Y \rightarrow X$, such that above each point in $X$, the complex structure of the torus fiber is literally equal to the value of the axio-dilaton $\tau$ at that point.
We are interested in backgrounds leading to $\mathcal{N}=1$ supersymmetry in $d=4$. This requires $X$ to be a K\"ahler threefold, and $Y$ to be a Calabi-Yau fourfold.

Although the torus directions of $Y$ are {\it per se} unphysical, one can still make sense of them through the dual M-theory picture:
Compactify M-theory on $Y$, whereby one fiber circle is the M-theory circle. Then, by a fiber-wise T-duality on the other circle $S_T^1$, one obtains IIB on $X \times \mathbb{R}^3 \ltimes \tilde S^1_T$, with $\tilde S^1_T$ is fibered over $X$. By shrinking the torus fiber, one recovers a Poincar\'e invariant $4d$ IIB vacuum with non-constant $\tau$.

This last chain of dualities relates a Euclidean D3-brane (E3-brane) on an internal four-cycle, or divisor $D$ of $X$, to a Euclidean M5-brane on a six-cycle $\hat D\subset Y$. $\hat D$ is a torus-fibration over $D$, i.e. $\hat D=\pi^{-1}(D)$.


In \cite{Witten:1996bn}, Witten found that the fermionic zero-modes of an M5-instanton on a six-cycle $\hat{D}$ of a CY fourfold $Y$ are counted by the Hodge numbers $h^{0,k}(\hat D)$. A necessary criterion for generation of a superpotential is that the holomorphic Euler characteristic equal one, i.e.\ $\sum_k (-1)^k\,h^{0,k}(\hat D)=1$, whereas the strong criterion requires that all such numbers vanish except for $h^{0,0}(\hat D)=1$.

While this criterion has become the `thumb rule' of instantons, it is often impractical from the IIB perspective, {\it especially} if worldvolume fluxes are to be included. In \cite{Bianchi:2011qh}, we recast the zero-modes of the M5-brane in terms of E3-brane data. Given a IIB background with varying axio-dilaton $\tau$, define the following one-form:
\begin{equation} \label{Qconnection}
Q_{\it 1}=\frac{\ii}{2}\frac{\d(\tau+\bar\tau)}{\tau-\bar\tau}=\frac{1}{2}\, e^\phi \, F_{\it 1}=\frac{\ii}2(\bar \partial \phi-\partial\phi)\,.
\end{equation}
This object can be regarded as the connection of a  $U(1)$ bundle, henceforth referred to as $U(1)_Q$,  defined so that the transition function acting on a typical section is a phase $e^{\ii\,{\rm arg}(c\,\tau + d)}$, whenever the background of $\tau$ undergoes an S-duality of the form (\ref{dualtau}).
From this $U(1)_Q$ bundle, one can construct a covariant derivative $\d_Q \equiv \d -\ii q\, Q_{\it 1}$, and an associated holomorphic line bundle $\mathcal{L}_Q$. By supersymmetry, this line bundle is directly related to the canonical bundle of the compactification threefold as
$
\mathcal{L}_Q \equiv K_{X}^{-1}.$

Our analysis shows that an E3-instanton in a generic $\tau$-background has left-handed and right-handed fermionic zero-modes given by the following (twisted) cohomologies:

\begin{table}[h!]
\begin{center}
\begin{eqnarray}\label{sheafcoho}
&&\begin{array}{c|c} \text{l.h.\ zero modes} &\text{cohomology group}\\ \hline
\lambda^\alpha_{\rm z.m.}&  H_{\del}^{0,0}(D)\\
\psi^\alpha_{\rm z.m.} & H_{\delbar}^{0,1}(D)\\
\rho^\alpha_{\rm z.m.} & H_{\del}^{2,0}(D)
\end{array} \qquad
\begin{array}{c|c} \text{r.h.\ zero modes} &\text{cohomology group}\\ \hline
\tilde\lambda^{\dot\alpha}_{\rm z.m.}&  H^0(D, \call_Q^{-1})\\
\tilde\psi^{\dot\alpha}_{\rm z.m.} &  H^{1}(D,\bar\call^{-1}_Q)\\
\tilde\rho^{\dot\alpha}_{\rm z.m.} & H^2(D, \mathcal{L}_Q^{-1})\end{array} \nonumber
\end{eqnarray}
\end{center}
\end{table}
\vskip 25mm
Here, we are displaying the zero modes of the appropriately topologically twisted worldvolume fermions $\lambda^\alpha, \psi^\alpha, \rho^\alpha, \tilde \lambda^{\dot \alpha}$, $\tilde \psi^{\dot \alpha}, \tilde \rho^{\dot \alpha}$, which are $(0,k)$ or $(k,0)$ forms on $D$  carrying also  (anti)chiral spinorial indices associated to the four non-compact  flat directions.
It can be shown that these E3-instanton modes are related to the M5-instanton fermionic zero-modes found in \cite{Witten:1996bn} as described by figure \ref{E3-M5}.
\begin{figure}[h]
\begin{center}
\begin{tikzpicture}[scale=1, node distance = 2cm, auto, inner sep=.1mm, text width=1.5cm, text centered]
\node (02m) at (0,.25) {$h^{2}_Q({\rm E3})$} ;
\node (02p) at (0,-.3) {$h^{2}({\rm E3})$};
\node (01m) at (0,-1.25) {$h^{1}_Q({\rm E3})$} ;
\node (01p) at (0,-1.8) {$h^{1}({\rm E3})$};
\node (00m) at (0,-2.75) {$h^{0}_Q({\rm E3})$} ;
\node (00p) at (0,-3.3) {$h^{0}({\rm E3})$};
\node[text width=4cm] at (-3,.25){\blue{\small geom.\ mod.}};
\node[text width=4cm] at (-3,-.3){\blue{\small tw.\ geom.\ mod.}};
\node[text width=4cm] at (-3,-1.25) {\blue{\small tw.\ Wilson lines}};
\node[text width=4cm] at (-3,-1.8) {\blue{\small Wilson lines}};
\node[text width=4cm] at (-3,-2.75) {\blue{\small $\tilde \lambda_{\rm z.m.}^{\dot\alpha}$}};
\node[text width=4cm] at (-3,-3.3) {\blue{\small $\lambda_{\rm z.m.}^{\alpha}$}};
\node (03) at (4,1.3) {$h^{3}({\rm M5})$};
\node (02) at (4,-.2) {$h^{2}({\rm M5})$};
\node (01) at (4,-1.7) {$h^{1}({\rm M5})$};
\node (00) at (4,-3.2) {$h^{0}({\rm M5})$};
\path [line] (00p) -- (00);
\path [line] (00m) -- (01);
\path [line] (01p) -- (01);
\path [line] (01m) -- (02);
\path [line] (02p) -- (02);
\path [line] (02m) -- (03);e3

\end{tikzpicture}
\end{center}
\caption{\small Schematic description of the relation between E3 and M5 fermionic zero modes. Here $h^k({\rm E3})\equiv h^{0,k}_{\delbar}(D)$,  $h^k_Q({\rm E3})\equiv h^k_Q(D) = \dim H^k(D, \call^{-1}_Q)$ and $h^{k}({\rm M5})\equiv\dim H^{0,k}_{\delbar}(\hat D)$. The zero modes $\lambda^\alpha_{\rm z.m.}$ and $\tilde\lambda_{\rm z.m.}^{\dot\alpha}$ correspond to the universal zero modes crucially studied in \cite{Bianchi:2007fx}, often denoted by $\theta^\alpha$ and $\bar\tau^{\dot\alpha}$ respectively, of  D-brane instantons in orientifold vacua.}\label{E3-M5}
\end{figure}

One can also repeat the anomaly argument of \cite{Witten:1996bn} to obtain the following sufficient condition for the generation of a superpotential: $\chi_{\rm E3}=1$, with $\chi_{\rm E3}:=\chi(D,\calo_D)-\chi(D,\call_Q^{-1})\equiv\sum_k(-)^n[h^{0,k}({\rm E3})-h^{0,k}_Q(\rm E3)]$.  By looking at figure \ref{E3-M5} it is easy to see that $\chi_{\rm E3}\equiv \chi(\hat D,\calo_{\hat D})$. Hence $\chi_{\rm E3}=1$ reproduces Witten's weak criterion in M-theory \cite{Witten:1996bn}.

In order to make contact with perturbative IIB string theory, one must define the double cover $\tilde X$ branched over a divisor locus of class $\call^2_Q\simeq K_X^{-2}$. This space is a Calabi-Yau threefold, with an involution of O7/O3 type. In this language, the cohomologies of the E3-brane split up into even and odd eigenspaces under the involution. Defining the double cover $\tilde D$ of the E3-brane on $D$, one can then see that our zero-modes map to perturbative IIB as follows:
\begin{equation}
\hskip 25mm h^{0,k}(D) \rightarrow h^{0,k}_+(\tilde D)\,, \quad h^{k}_Q(D) \rightarrow h^{0,k}_-(\tilde D)\,.
\end{equation}
Hence, in the orientifold limit we recover the equivariant cohomologies of E3-divisors seen in \cite{Blumenhagen:2010ja}.

\section{Magnetized instantons}
Our classification of the zero-modes in terms of $\mathcal{L}_Q$-valued forms clarifies how to think of these objects in a very general way: On the one hand, it allows one to count them directly in IIB away from the weak coupling limit, on the other hand, it relate the zero-modes found in orientifold setups directly to the zero-modes in M-theory.

However, the most interesting application of this new language is the possibility to include E3-worldvolume flux. This is particularly difficult to understand in the M-theory picture because it uplifts to the problematic self-dual three-form on the M5-brane. In IIB, however the world-volume flux can be understood as a $(1,1)$-form Poincar\'e dual to a sum of holomorphic curves on the E3-divisor. This allows for a more viable approach. In this section, we display the upshot of our general analysis in \cite{Bianchi:2011qh}, which gives the fermionic action and zero-modes for a magnetized E3-brane in the presence of a generic $\tau$ background.

The worldvolume DBI flux on the E3-brane transforms as a doublet under S-duality:
\begin{equation}
\left(\begin{array}{c} \calf^D \\
\calf\end{array}\right)\rightarrow \left(\begin{array}{cc} a & b \\ c &
d\end{array}\right)\left(\begin{array}{c} \calf^D \\ \calf\end{array}\right)\,,
\end{equation}
where $\calf = 2 \pi \alpha' F_{E3}-i^*B_2$ is the `electric' flux, and ${\cal F}^D$ is the `magnetic' dual flux. The BPS equations of holomorphy and anti-self-duality impose that:
\begin{equation}
\calf^{0,2} = \calf^{2,0}=0\,, \qquad \tau\,\calf=\calf^D\,.
\end{equation}
From this one can deduce that $\calf$ is a 2-form transforming as a section of $\call_Q$ under S-duality:
\begin{equation}
\bar \partial_Q(e^{-\phi}\,\calf) = 0 \sim \bar \partial (\calf) = 0 \qquad \iff \calf \in H^{1}(D, T^*_D \otimes \call_Q)\,.
\end{equation}
In IIB orientifold language, this corresponds to saying that $\calf \in H^{1,1}_-$.

In order to arrive at our final result, we will introduce two more objects rather briefly, referring to \cite{Bianchi:2011qh} for details. The first is $\Omega$, a holomorphic $(3,0)$-form that takes values in $\call_Q$, i.e. $\Omega \in K_{X} \otimes \call_Q$. In the case of constant $\tau$ and trivial $\call_Q\simeq K_X^{-1}$, it reduces to the usual CY threeform. The second set of objects are composite tensors built with the extrinsic curvature $\calk^m{}_{ab}$ of the E3-divisor in $X$:
\begin{eqnarray}
(\cals_\calf)_{\bar\imath\bar\jmath}{}^{uv}&=&-e^{-\phi}\calk^m{}_{\bar t[\bar\imath}\calf_{\bar\jmath]}{}^{\bar t}
\bar\Omega_{m}{}^{uv}\cr
(\tilde\cals_\calf)_{ij}{}^{\bar u\bar v}&=&-e^{-\phi}\calk^m{}_{t[i}\calf_{j]}{}^{t}
\Omega_{m}{}^{\bar u\bar v}\,.
\end{eqnarray}
Starting from the general action of \cite{dirac}, we obtain the following one for the topologically-twisted fermions:
\begin{eqnarray}\label{actionflux2}
S_{\rm F}&=&\int_{D} \big(\psi\wedge
*\del\lambda-\tilde\psi\wedge *\delbar_Q\tilde\lambda-\rho\wedge*\delbar\psi+\tilde\rho\wedge*\del_Q\tilde\psi\big)\cr
&&\quad\qquad+\int_{D}\sqrt{\det h} \,\big(\rho\cdot \cals_\calf\cdot \rho -\tilde\rho\cdot\tilde\cals_\calf\cdot \tilde\rho\big)
\end{eqnarray}
The first line reproduces the expected equations for the zero-modes in the fluxless case. The second line, on the other hand, shows that $\rho$ and $\tilde \rho$ acquire masses through the presence of background DBI fluxes and can then be lifted by them.
The $\cals$ operators can be regarded as cohomology maps:
\begin{eqnarray}
\cals_\calf&:& H^{2,0}_\del(D)\rightarrow H^{0,2}_{\delbar}(D)\cr
\tilde\cals_\calf &:& H^{0,2}_{\delbar}(D, {\cal L}_Q^{-1})\rightarrow H^{2,0}_{\del}(D,{\cal L}_Q)\,.
\end{eqnarray}
In this language, the fermionic zero-modes for an E3-instanton can be recast into the following final form:
\begin{eqnarray}\label{sheafcoho2}
\begin{array}{c|c} \text{\lhdot\ zero modes} &\text{vector space}\\ \hline
\lambda^\alpha_{\rm z.m.}& H^0(D,\bar\calo_D)\\
\psi^\alpha_{\rm z.m.} &  H^{1}(D,\calo_D)\\
\rho^\alpha_{\rm z.m.} &  \ker \cals_\calf \subset   H^{2}(D,\bar\calo_D)
\end{array}\qquad \begin{array}{c|c} \text{\rhdot\ zero modes} &\text{vector space}\\ \hline
\tilde\lambda^{\dot\alpha}_{\rm z.m.}&  H^0(D, \call_Q^{-1})\\
\tilde\psi^{\dot\alpha}_{\rm z.m.} &  H^{1}(D,\bar\call^{-1}_Q)\\
\tilde\rho^{\dot\alpha}_{\rm z.m.} & \ker \tilde\cals_\calf\subset   H^{2}(D, {\cal L}_Q^{-1})\end{array}
\end{eqnarray}
Importantly, one can show  \cite{Bianchi:2011qh} that  this flux-induced lifting of the $\rho$ and $\tilde\rho$ zero modes is in precise correspondence with the flux-induced  lifting of the geometric zero-modes superpartners, which was discussed above. In the following section we will exploit this geometric characterization in order to look for E3-branes wrapping non-rigid divisors but nevertheless contributing to the superpotential.

\section{Example}
In this section, we show that all of our results can have a fairly simple incarnation in algebraic geometry. In order to freeze an instanton via worldvolume fluxes one has to find rigid holomorphic curves in the threefold and impose that the E3-divisor entirely contain them.

The setting is a simple one-modulus K\"ahler threefold: $X=\mathbb{P}^3$. We will treat it in the perturbative IIB limit, where we will use its CY double-cover $\tilde X$. It is defined as the degree eight hypersurface in the weighted projective fourfold $\mathbb{P}^4_{1 1 1 1 4}$, with homogeneous coordinates $[z_1: \ldots : z_4: \xi]$. Our hypersurface equation will be:
\begin{equation} \label{hypersurf}
\xi^2+z_1^8+ \ldots + z_4^8+\psi^2\,P_4(z_i)^2 =0\,, \quad \text{with } \psi \in \mathbb{C}\,,
\end{equation}
and $P_4$ is some generic polynomial of homogeneous degree four. The orientifold involution $\sigma$ acts on $\tilde X$ as a reflection, $\sigma: \xi \rightarrow -\xi$. This CY threefold has one K\"ahler modulus and  contains no rigid divisors, hence it is traditionally ruled out for generating non-perturbative superpotentials \cite{Robbins:2004hx}.

The simplest divisor $D$ (here,  for simplicity, we use the same symbol for $D$ and $\tilde D$) on which an E3 can be wrapped is a divisor given by the vanishing locus of a degree one polynomial $P_D$:
\begin{equation} \label{divisor}
P_D: a_1 z_1+ \ldots + a_4 z_4=0\,,
\end{equation}
where the $a_i$ are arbitrary complex numbers. Since an overall rescaling of $P_D$ doesn't change the actual divisor, the number of geometric deformations of $D$ is $h^{0,2}_-=3$. On the other hand, the divisor is automatically invariant under $\sigma$ and then it has $h^{0,2}_+=0$.\footnote{Note that $h^{0,2}_-$ corresponds to {\it invariant} geometric deformations. Since the $\Omega$ 3-form is itself odd, it maps even deformations to odd $(0,2)$-forms.} In our language, this instanton has, beside the two omnipresent $\lambda_{\rm z.m.}^\alpha$ zero-modes, 3 zero-modes $\tilde \rho_{\rm z.m.}^\alpha$ that spoil superpotential generation.

Our claim is that all three zero-modes can be lifted by an appropriate flux. A holomorphic flux on a divisor is Poincar\'e dual to a linear combination of Riemann surfaces. CY threefolds are known to contain huge amounts of rigid spheres that are usually counted by topological invariants. In this particular case, there are $29504$ isolated degree one $\mathbb{P}^1$'s. We can fix the  $a_i$'s  in such a way that $D$ contains some of these curves and choosing the flux on $D$ to be dual to such two-cycles. These $\mathbb{P}^1$'s being isolated in ${\tilde X}$, if the divisor moves the two-cycle dual to the flux will continue to exist, but will no longer be holomorphic. Dually, $\calf^{0,2}$ will cease to be vanishing, hence realizing our zero modes lifting mechanism.

Define the curve $C_1$  as the complete intersection of three equations in $\mathbb{P}^4_{1 1 1 1 4}$:
\begin{eqnarray}
C_1: \quad z_1 &=& \eta\,z_2 \quad \cap \quad z_3 = \tilde \eta\, z_4 \quad \cap \quad \xi = \psi\,P_4 \qquad \subset \quad \mathbb{P}^4_{1 1 1 1 4}\,,
\end{eqnarray}
where $\eta^8= \tilde \eta^8 = -1$. Its orientifold image $C_1'$ is defined in the same way up to the sign change $\xi\rightarrow -\xi$.

Satisfying these three equations automatically implies that \eqref{hypersurf} is zero. Hence, both $C_1$ and $C'_1$ lie in $\tilde X$. It can be easily shown that they are $\mathbb{P}^1$'s, and with more work, that they are rigid.
If we define $\calf$ via Poincar\'e duality as $\calf \simeq {\rm PD}_D(C_1-C_1')$, this flux will be odd under $\sigma$, as required.\footnote{This flux also needs a half-integrally quantized component \cite{Freed:1999vc}. This issue is addressed in detail in \cite{Bianchi:2011qh}.} In order for the curves to be holomorphic contained in $D$, the $a_i$'s in $P_D$ \eqref{divisor} must satisfy $a_1\,\eta+a_2=0$ and $a_3\,\tilde\eta+ a_4=0$.
This kills off two out of three $\tilde \rho$ zero-modes. The E3-brane now has the following shape: $P_D (C_1)= a_1\,(z_1-\eta\,z_2) + a_3\,(z_3-\tilde \eta\,z_4) = 0$,
hence with one geometric modulus. The latter can be easily fixed by adding another curve/image-curve pair to the flux:
\begin{equation}
C_2, C_2': \quad z_1 = \eta\,z_4 \quad \cap \quad z_3 = \tilde \eta\, z_2 \quad \cap \quad \xi = \pm\psi\,P_4 \qquad \subset \quad \mathbb{P}^4_{1 1 1 1 4}
\end{equation}
which imposes the further restrictions $a_1\,\eta +a_4 = 0$ and $a_3\,\tilde \eta+a_2=0$.

Hence, a flux $\calf \simeq {\rm PD}_D\left(C_1+C_2-C_1'-C_2' \right)$ completely freezes the E3-brane to be given by:
\begin{equation}
P_D=a( z_1-\eta\,z_2+\tfrac{\eta}{\tilde \eta}\,z_3-\eta\,z_4)  = 0\,.
\end{equation}
The results of our previous sections then imply that this magnetized E3-brane has exactly the two fermionic zero-modes and can generate a non-perturbative superpotential.\footnote{One must also be aware of the so-called `charged' zero-modes present whenever the E3 intersects a D7-brane, \cite{Blumenhagen:2007sm}. However, our fluxes neither aggravate nor improve this issue.}

\section{Conclusion}
We have recast Witten's analysis of M5-instanton zero-modes in IIB language, in order to treat E3-instantons in non-trivial backgrounds for $\tau$. This entails viewing the fermionic zero-modes as forms on the E3-brane taking values in a line bundle $\call_Q$, that is dual to the canonical bundle of the threefold by supersymmetry. The analysis confirms the results of \cite{Blumenhagen:2010ja} in the orientifold limit.

We generalized the analysis to include DBI fluxes on the worldvolume of the E3-brane. We were able to write an $SL(2, \mathbb{Z})$-covariant action for the fermions, and found that the fluxes indeed create masses for the zero-modes corresponding to geometric deformations of the E3-divisor.

All of this machinery was put to use to illustrate that even a simple one-modulus, non-toroidal CY threefold with an O7 plane, the $\mathbb{P}^4_{1 1 1 1 4}[8]$, contains a divisor that generates a non-perturbative superpotential, after being rigidified by a flux dual to a rigid curve.

The fact that we used a very simple form of the hypersurface equation is irrelevant. CY threefolds typically contain thousands of rigid holomorphically embedded $\mathbb{P}^1$'s, counted by the Gromov-Witten invariants. Hence, we expect such rigidifying fluxes to be a generic feature in the landscape.

One should also note that instantons are not objects that can be `put in' at will, but must be summed over, and so must their worldvolume fluxes. Hence, any IIB compactification of this type is bound to receive many more corrections to the superpotential due to instanton effects than previously believed. We hope that this work has shed some new light on the landscape problem, and will inspire new explorations for models of moduli stabilization.

\subsection*{Acknowledgements}
A. C. is a Research Associate of the Fonds de la Recherche Scientifique F.N.R.S. (Belgium). The work of M.~B.\ and L.~M.\ was partially supported by the ERC Advanced Grant n.226455 "Superfields", by the Italian MIUR-PRIN contract 2009KHZKRX-007,
by the NATO grant PST.CLG.978785.

\end{document}